\documentclass{aa}

\usepackage{txfonts}

\usepackage{graphicx}

\usepackage{natbib}
\bibpunct{(}{)}{;}{a}{}{,} 

\def\solar {\ifmmode_{\mathord\odot} \else $_{\mathord\odot}$\fi}
\def\Msol {\ifmmode {\,{\it M}\solar} \else $\,M$\solar\fi}     
\def\Rsol {\ifmmode {\,{\it R}\solar} \else $\,R$\solar\fi}     
\def\Lsol {\ifmmode {\,{\it L}\solar} \else $\,L$\solar\fi}     
\newcommand{\Mjup}{M$_{Jup}$}
\newcommand{\Mearth}{M$_{Earth}$}

\begin{document}

\title{The HARPS search for southern extra-solar planets
       \thanks{Based on observations made with the HARPS instrument on the
         ESO 3.6-m telescope at La Silla Observatory under programme ID
         072.C-0488}
       }
\subtitle{XIV. Gl~176b, a super-Earth
rather than a Neptune, and at a different period}

\author{
T. Forveille \inst{1}
\and X. Bonfils \inst{2,1,3}
\and X. Delfosse \inst{1}
\and M. Gillon \inst{4}
\and S. Udry \inst{4}
\and F. Bouchy \inst{5}
\and C. Lovis \inst{4}
\and M. Mayor \inst{4}
\and F. Pepe \inst{4}
\and C. Perrier \inst{1}
\and D. Queloz \inst{4}
\and N. Santos \inst{2}
\and J.-L. Bertaux \inst{6}
}

\offprints{T. Forveille, \email{Thierry.Forveille.ujf-grenoble.fr}}

\institute{Laboratoire d'Astrophysique de Grenoble,
               Observatoire de Grenoble,
               Universit\'e Joseph Fourier,
               CNRS, UMR 571
               Grenoble,
               France
\and
               Centro de Astrof{\'\i}sica, Universidade do Porto, 
               Rua das Estrelas, 
               4150-762 Porto, Portugal
\and
               Centro de Astronomia e Astrof{\'\i}sica da 
                    Universidade de Lisboa,
               Observat\'orio Astron\'omico de Lisboa, 
               Tapada da Ajuda, 
               1349-018 Lisboa, Portugal
\and
               Observatoire de Gen\`eve, 
               Universit\'e de Gen\`eve, 
               51 ch. des Maillettes, 
               1290 Sauverny,
               Switzerland
\and  
	       Institut d'Astrophysique de Paris, 
               CNRS, Universit\'e Pierre et Marie Curie, 
               98bis Bd Arago, 75014 Paris, France
\and
	       Service d'A\'eronomie du CNRS, BP 3, 
               91371 Verri\`eres-le-Buisson, France	
}

\abstract
{A 10.24 days Neptune-mass planet 
was recently announced to orbit the nearby M2 dwarf Gl~176, based
on 28 radial velocities measured with the HRS spectrograph on 
the Hobby-Heberly Telescope.}
{ }
{We obtained 57 
radial velocities of Gl~176 with the ESO 3.6m telescope
and the HARPS spectrograph, which is known for its sub-m/s stability.
The median photon-noise standard error of our measurements is 
1.1~m/s, significantly lower than the 4.7~m/s of the 
HET velocities, and the 4~years period over which they were
obtained has much overlap with the epochs of the HET measurements.
}
{The HARPS measurements show no evidence for a signal at the
period of the putative HET planet, suggesting that its detection was
spurious. We do find, on the other hand, strong evidence for
a lower mass 8.4~M$_{Earth}$ planet, in a quasi-circular orbit 
and at the different period of 8.78~days.
The host star has moderate magnetic activity and rotates on 
a 39-days period, which we confirm through modulation of
both contemporaneous photometry and chromospheric indices. 
We detect that period as well in the radial
velocities, but it is well removed from the orbital period and
no cause for confusion.
}
{This new detection of a super-Earth 
(2~\Mearth$<$~M~sin(i)~$<$~10~\Mearth)
around an M dwarf adds
to the growing evidence that such planets are common around
very low mass stars: a third of the 20 known planets with 
M~sin(i)~$<$~0.1~\Mjup and 3 of the 7 known planets with 
M~sin(i)~$<$~10~\Mearth orbit an M dwarf, in contrast to
just 4 of the $\sim$300 known Jupiter-mass planets.}

\date{}


\titlerunning{A super-Earth planet around the nearby M dwarf Gl~176}
\authorrunning{Forveille et al.}

\maketitle

\section{Introduction}

Of the $\sim$250 planetary systems currently known from radial 
velocity monitoring, 
just half a dozen 
are centered around M dwarfs (M$<$0.6\Msol)
\footnote{http://exoplanet.eu/catalog-RV.php}.
This in part reflects a selection bias, since an order of
magnitude fewer faint M dwarfs are searched for planets
than brighter solar-type stars are, but M dwarfs also 
seem to genuinely have fewer massive planets 
($\sim$\Mjup) than the more massive solar-type stars do
\citep{Bonfils2006,Johnson2007}. They seem on the other 
hand  \citep{Bonfils2006} to have larger numbers of the 
harder to detect Neptune-mass and super-Earth planets:
a third of the $\sim$20~planets with M~sin(i)~$<$~0.1\Mjup 
known to date orbit an M dwarf, in
spite of solar-type stars outnumbering those by an order of
magnitude in planet search samples. As a consequence of their 
small overall number, each individual M-dwarf planetary system 
still plays a significant role in defining these emerging 
statistical properties.

Very recently, \citet{Endl2008} announced the discovery of a 
planet with a minimum mass of M~sin(i)~=~25~M$_{Earth}$ 
in a 10.24~days orbit around a nearby M2.5 dwarf. Gl~176 
(also HD~285968, HIP~21932, LHS~196) is a V=9.97 
\citep{Upgren1974} member of the immediate solar neighborhood 
\citep[par=106.2$\pm$2.5~mas, d~=~9.4~pc,][]{Hipparcos1997}. The 
2MASS photometry \citep{2MASS2006} and the parallax result in
an absolute magnitude of M$_{Ks}$~=~5.74, and together with the
K-band mass-luminosity relation of \citet{Delfosse2000}
in a mass of 0.50\Msol. Based on the \citet{Bonfils2005} photometric
metallicity calibration, [Fe/H] is -0.1$\pm$0.2 and therefore
solar within its uncertainty.  
\begin{table}
\centering
\caption{
\label{table:stellar}
Observed and inferred stellar parameters for Gl~176}
\begin{tabular}{l@{}lc}
\hline
 \multicolumn{2}{l}{\bf Parameter}
& \multicolumn{1}{c}{\bf Gl~176} \\
\hline
Spectral Type   &                & M2V\\
V                       &               & $9.97 \pm 0.03$ \\
$\pi$           &[mas]          & $106.16 \pm 2.51$ \\
Distance                &[pc]           & $9.42 \pm 0.22  $\\
$M_V$           &               & $10.10 \pm 0.06$ \\
K                       &               & $5.607 \pm 0.034$\\
$M_K$           &               & $5.74 \pm 0.06 $\\
$L_\star$       & [$\mathrm{L_\odot}$]          &  $0.022$\\
$L_x/L_{bol}$   &               &  $3.5.10^{-5}$\\
$v\sin i$               & [km\,s$^{-1}$] & $ \lesssim 0.8 $ \\
$[Fe/H]$                &               & $ -0.1 \pm 0.2 $\\
$M_\star$       & [$\Msol$]             & $ 0.50 $\\
\hline
\end{tabular}
\end{table}

We have independently been monitoring the radial velocity of
Gl~176 using the HARPS spectrograph on the ESO 
3.6-m telescope, over a period which largely overlaps
the epochs of the \citet{Endl2008} observations. Section 2. 
describes those independent measurements and concludes 
that they do not confirm the 10.24~days planet. Section 3.
takes a closer look at those mesurements and finds
that they contain two coherent signals, with periods of
8.78 and 40.0~days. Section~4 discusses differential 
photometry and variation of chromospheric indices, to 
conclude that the 40~days signal reflects the stellar
rotation period. The 8.78~days period on the other hand
is due to a bona-fide planet, with a minimum mass of
only 8.4~\Mearth. Section~5 concludes with a brief 
discussion of the new planet.
 

\section{HARPS Doppler measurements and orbital analysis}
We observed Gl~176 with HARPS (High Accuracy Radial velocity 
Planet Searcher) as part of the guaranteed-time program of the
instrument consortium. HARPS is a
high-resolution (R~=~115 000) fiber-fed echelle spectrograph, optimised
for planet search programmes and asteroseismology. It is the most 
precise spectro-velocimeter to date, with a long-term instrumental
RV accuracy under 1~m\,s$^{-1}$ \citep{Mayor2003,Santos2004,Lovis2005}.
For ultimate radial velocity precision HARPS uses
simultaneous exposures of a thorium lamp through a calibration fiber.
When observing M dwarfs however, we rely instead on its very high
instrumental stability (nightly instrumental drifts $<$~1\,m~s$^{-1}$).
Most M dwarfs are too faint for us to reach the stability limit
of HARPS within realistic integration times, and dispensing with the
simultaneous thorium light produces much cleaner stellar spectra, suitable
for quantitative spectroscopic analyses.

For the V~=~9.97 Gl~176 we use 15~mn exposures, and the 
median S/N ratio of our 57 spectra is 60 per pixel at 550~nm. 
The radial velocities (Table~\ref{TableRV}, only available 
electronically) were obtained with the standard HARPS 
reduction pipeline, based on cross-correlation 
with a stellar mask and a precise nightly
wavelength calibration from ThAr spectra \citep{Lovis2007}.
They have a median internal error of only 1.1~m\,s$^{-1}$, 
which includes both the nightly zero-point calibration 
uncertainty ($\sim$~0.5~m\,s$^{-1}$)
and the photon noise, computed from the full Doppler
information content of the spectra \citep{Bouchy2001}.

\begin{figure}
\includegraphics[width=9cm,angle=0]{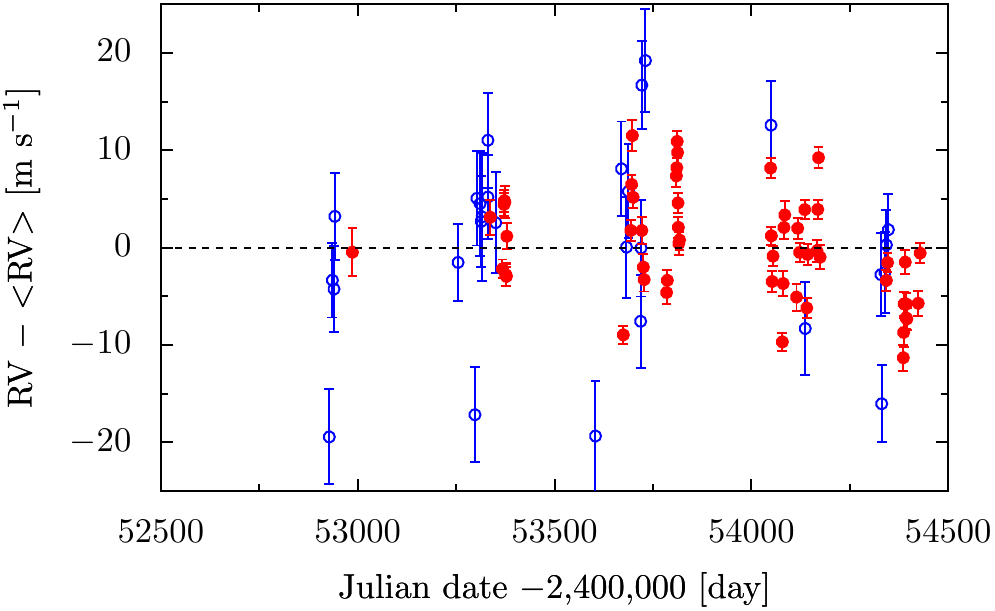}
\includegraphics[width=9cm,angle=0]{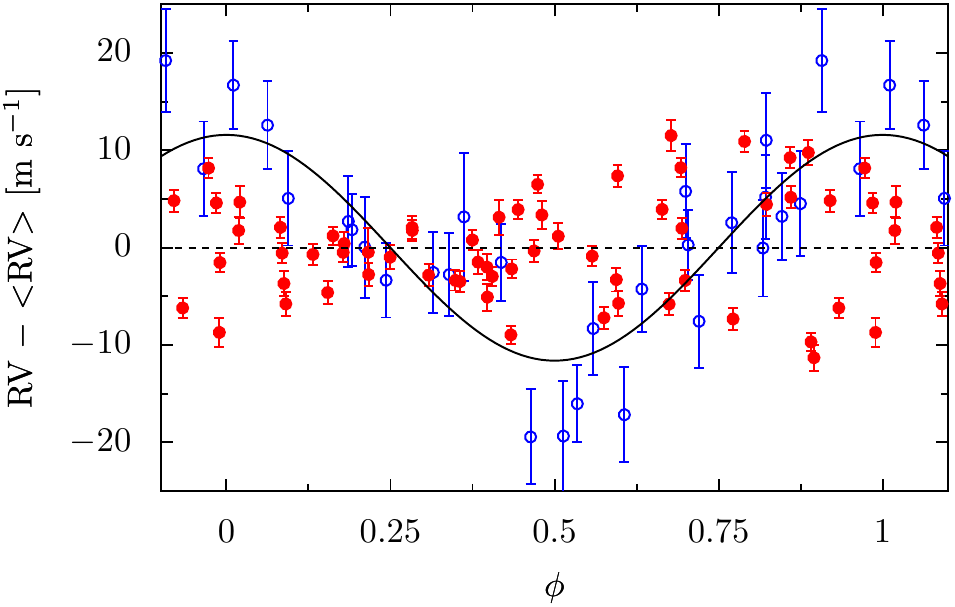}
\caption{Top panel: HARPS (red filled symbols) and \citet{Endl2008} 
(blue empty symbols) radial velocities of Gl~176 as a function of time,
overlaid with the prediction of the \citet{Endl2008} 
orbit. Bottom panel: HARPS radial velocities phased at the
10.24~days period of the \citet{Endl2008} orbit, overlaid 
with the radial velocity prediction for that orbit.
}
\label{Endl_Comparison}
\end{figure}

The computed velocities exhibit an rms dispersion of 5.3~m\,s$^{-1}$.
This is much above the 1~m\,s$^{-1}$ internal errors and significantly 
more than we observe for stars with similar chromospheric activity, but 
less than the $\sim$8~m\,s$^{-1}$ expected from the 
11.7~m\,s$^{-1}$ velocity amplitude  of the \citet{Endl2008} 
orbit. 
Fig.\ref{Endl_Comparison} confirms that the HARPS velocities are 
more tightly packed than both the HET measurements (top panel)
and the predictions of the \citet{Endl2008} orbit (lower panel). Its 
lower panel demonstrates that they do not phase on the \citet{Endl2008}
period, and we verified that the subset of the HARPS dataset
which overlaps the published HET measurements does not either.
Since any instrumental or astrophysical noise can only
increase the velocity dispersion, never decrease it,
the HARPS measurements set a $\sim$7.5~m\,s$^{-1}$ ceiling 
on the radial velocity amplitude of a Keplerian orbit (except 
for unrealistically high eccentricities). This forces us to 
conclude that the \citet{Endl2008} orbit must be spurious, though
we do not have a ready explanation for why. 

\section{\label{sect:orbit}Orbital analysis}
\begin{figure}
\centering
\includegraphics[width=0.9\linewidth]{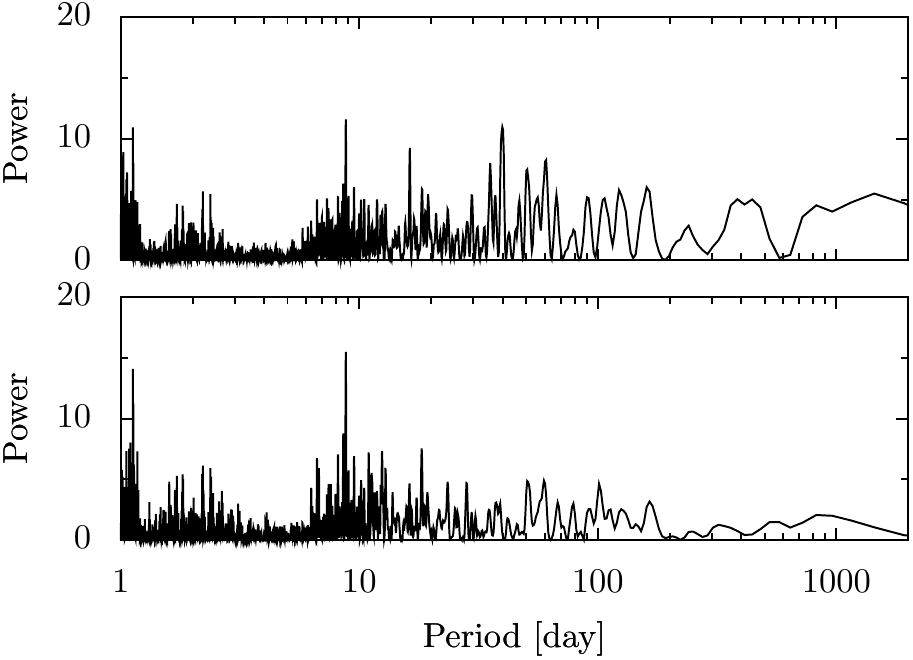}
       \caption{
	Lomb-Scargle periodogram of the raw HARPS radial velocities 
        (top panel), and of the velocities after subtraction of the
        40-day signal (bottom panel). }
       \label{Fig-Periodogram}
\end{figure}

Our radial velocity measurements do show coherent structure,
and a Lomb-Scargle periodogram \citep{Press1992} 
shows two narrow peaks around 8.8 and 40 days  
(Fig.~\ref{Fig-Periodogram}, top panel). 
The two peaks have similar 
false alarm probabilities of 0.1\%, and their spacing
is well removed from any significant feature in the 
window function. We therefore analysed them simultaneously
and searched for 2-planet Keplerian solutions with
{\it Stakanof} (Tamuz, in prep.), a program
which uses genetic algorithms to efficiently
explore the large parameter space of multi-planet
models. {\it Stakanof} robustly converged to a
2-keplerian solution with periods which match
the two periodogram peaks. Subtracting the longer
period signal from the velocities increases the
significance of the 8.8~day period in the periodogram 
(Fig.~\ref{Fig-Periodogram}, lower panel), further 
increasing our confidence that this signal is real. 
Subtracting the short period signal, on the other
 hand, produces a periodgram (not shown)
with a less convincing 40~days peak.

The 2-planet model describes our measurements well, but 
certainly not perfectly 
($\sigma$~=~2.5~m\,s$^{-1}$, $\sqrt{\bar{\chi}^2}~=~2.46$ 
per degree of freedom). A Lomb-Scargle periodogram 
of the residuals of this 2-planet solution however 
shows no significant peak, and the significant residuals 
therefore contain no immediate evidence for an additional 
component. 

\begin{table}
\caption{Orbital elements for the two-keplerian
orbital model of Gl\,176.}
\label{TableElements}
\centering
\begin{tabular}{c c c}
\hline\hline
\bf Element & \bf Value & \bf Standard error \\
\hline
$\gamma$ &  26.4105 km/s &  0.0004\\
\\
P$_1$ [days] &   8.7836    &    0.0054 \\
e$_1$  &   0.0          &   Fixed \\
om$_1$  [deg.] &   0.0 &  Fixed \\
T0$_1$  [jdb] &    2454399.79 &   0.33 \\
K1$_1$  [m/s] &          4.12     &    0.52 \\
\\
P$_2$  [days]  &        40.00 &    0.11\\
e$_2$        &         0.0 &    Fixed \\
om$_2$  [deg.] &       0.0 &   Fixed \\
T0$_2$  [jdb]  &    2454291.07 &   1.31 \\
K1$_2$  [m/s]  &           4.23 &      0.53 \\
\hline
\end{tabular}
\end{table}

\begin{figure}
\centering
\includegraphics[width=0.9\linewidth]{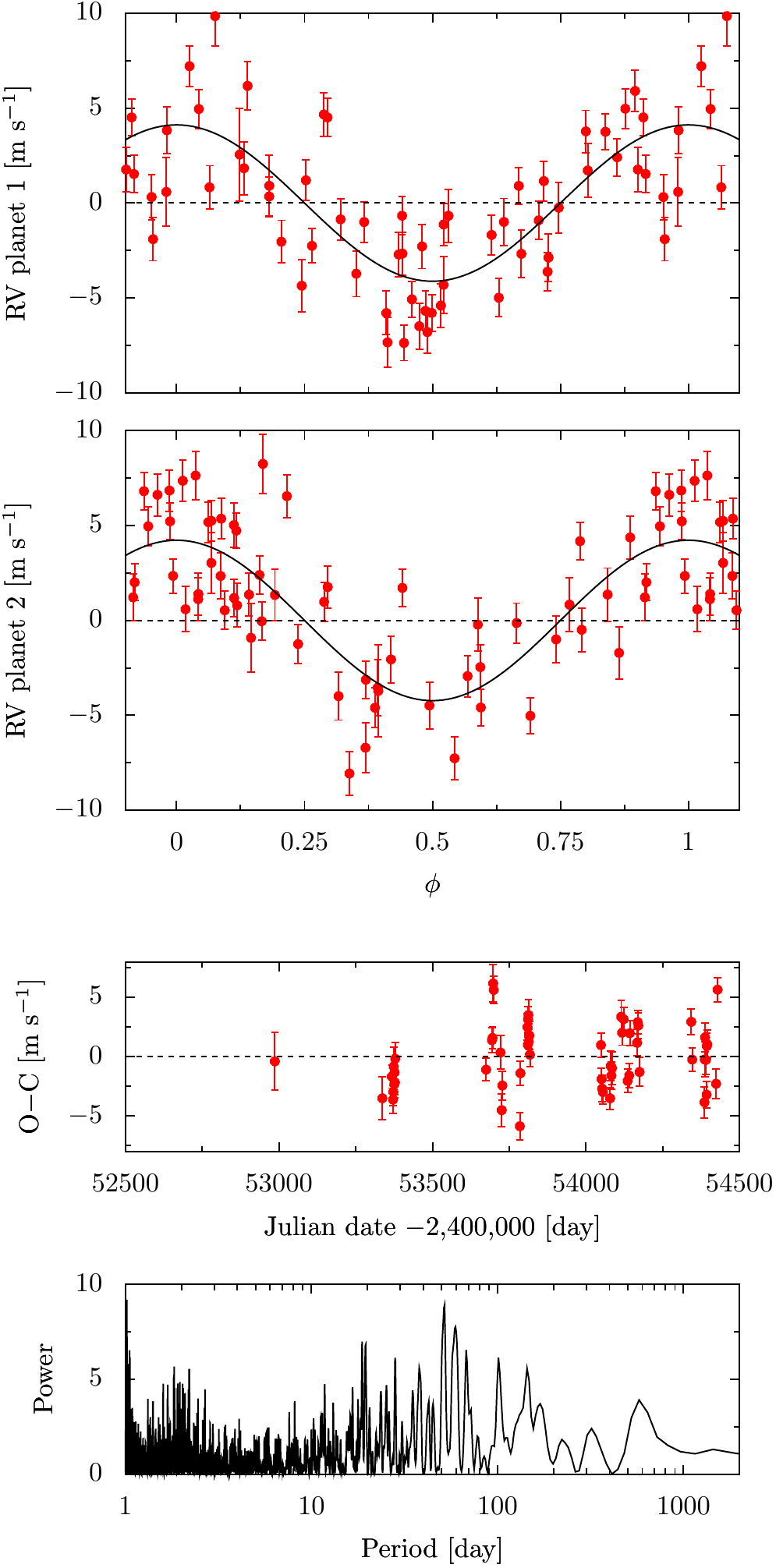}
       \caption{
 {\it Top two panels:} Radial velocity measurements phased to each
  of the two periods, after subtraction of the other component of 
  our best 2-planet model. 
{\it Third panel:} Residuals of the best 2-planet fit as a function 
  of time (O$-$C, Observed minus Computed).
{\it Bottom panel:} Lomb-Scargle periodogram of these residuals.
}
       \label{fig4}
\end{figure}

Both Keplerian signals have amplitudes of $\sim$4~m\,s$^{-1}$,
which with hindsight is well under the sensitivity 
limit of \citet{Endl2008}. Neither of their eccentricities is
significant, and we therefore adopt circular orbits as 
our preferred solution (Table~\ref{TableElements}, 
Fig.~\ref{fig4}); that choice does not affect any of our
conclusions. The inner and outer planets,
in a Keplerian interpretation of the radial velocity
variations, have minimum masses ($m \sin i$) of
8 and 14~\Mearth and projected semi-major axes 
of 0.066 and 0.18~AU. 


\section{\label{sect:activity}Activity analysis}
\begin{figure}
\centering
\includegraphics[width=0.9\linewidth]{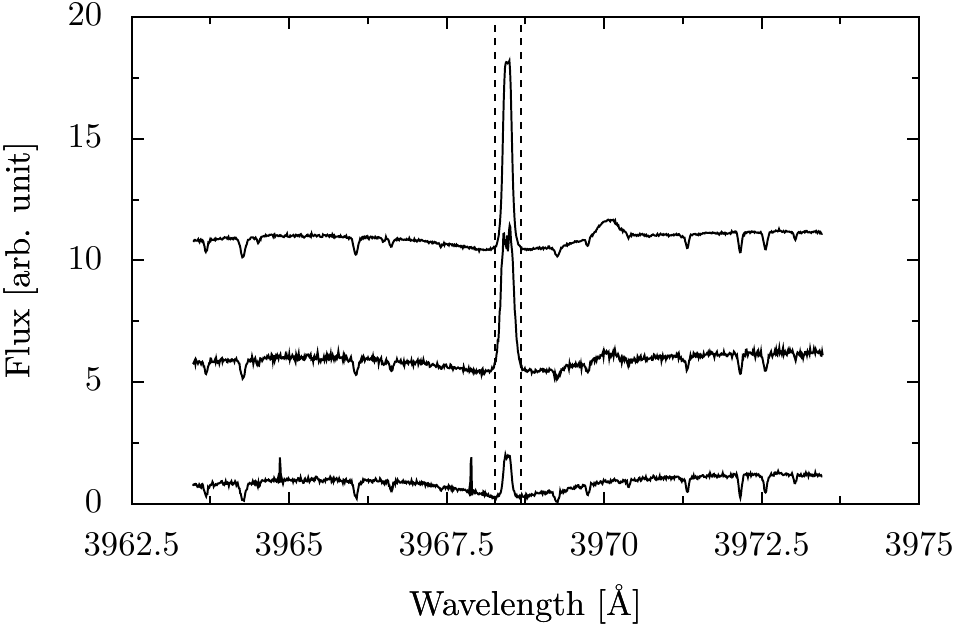}
       \caption{Emission reversal in the 
       \ion{Ca}{ii} H line in the average spectra of 
       \object{Gl~674} (M2V, top), \object{Gl~176} (M2.5V, middle),
       and \object{Gl 581} (M3V, bottom). Within our 100~M
       dwarfs sample, 
       \object{Gl 581} has one of the weakest \ion{Ca}{ii} 
       emission and illustrates a very quiet M dwarf.
       \object{Gl~674} and \object{Gl~176} have much stronger 
       emission and are both moderately active.
       }
       \label{Fig_CaII}
\end{figure}

Apparent Doppler shifts unfortunately do not always originate
in the gravitational pull of a companion, because stellar 
surface inhomogeneities, such as plages and spots, can break the 
balance between light emitted in the red-shifted and the
blue-shifted parts of a rotating star. These inhomogeneities 
then translate into rotationally modulated changes of 
both the shape and the centroid of spectral lines 
\citep[e.g.][]{Saar1997, Queloz2001}. The activity
level of  \object{Gl~176} is similar to that of
\object{Gl~674} (Fig.~\ref{Fig_CaII}), where a
spot is responsible for a 5~m\,s$^{-1}$ radial
velocity signal \citep{Bonfils2007}.

For well resolved rotational broadenings, correlated 
variation in the shape, parametrized by the line bisector, 
and in the centroid, 
provide an excellent diagnostic of such apparent velocity
variations. We however measure from our \object{Gl~176} 
spectra a rotational velocity of 
$v\sin i \lesssim 0.8~\mathrm{km\,s^{-1}}$.
This small rotation velocity removes
much of the usual power of the bisector test,
since the bissector span scales with a much
higher power of $v\sin i$ than the centroid
\citep[][]{Saar1997,Bonfils2007}. 

Spots fortunately also produce flux variations, and 
they typically impact  spectral indices, whether designed 
to probe the chromosphere (to
which photospheric spots have strong magnetic connections), or
the photosphere (because spots have cooler spectra). 
We therefore investigated 
the magnetic activity of \object{Gl~176} through 
photometric observations 
(\S\ref{subsect:phot}) and detailed examination of the 
chromospheric features in the clean HARPS spectra 
(\S\ref{subsect:spec}).

\subsection{\label{subsect:phot}Photometric variability}

\begin{figure}
\centering
\includegraphics[width=0.9\linewidth]{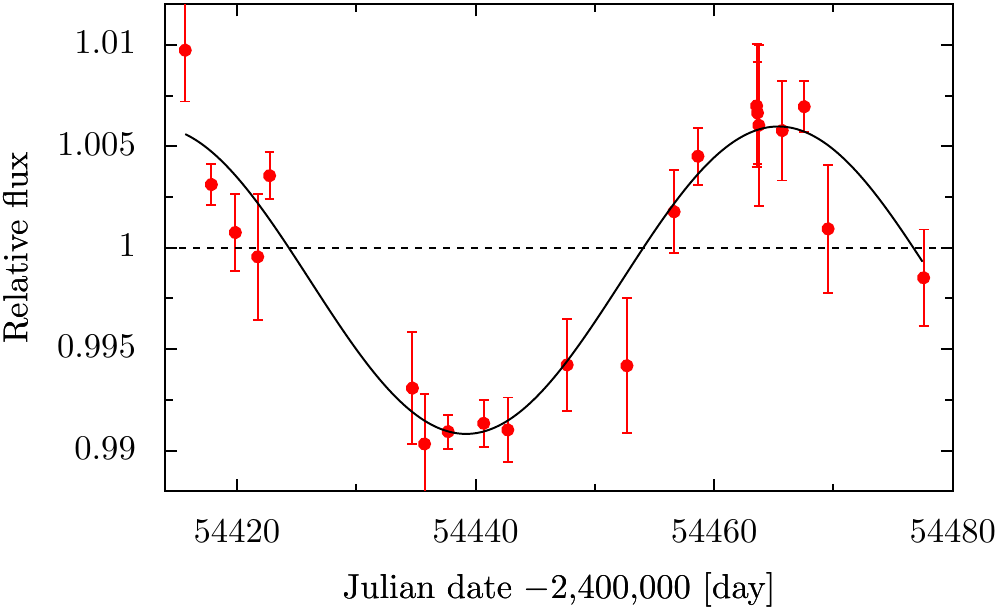}
       \caption{{\it Upper panel:} Differential
       photometry of \object{Gl~176} as a
       function of time. The star clearly varies 
       on a 40-50~days time scale with a $\sim$1.3\% peak 
       to peak amplitude 
       }
       \label{fig:phot1}
\end{figure}
We obtained photometric measurements with
the EulerCAM CCD camera of the Euler Telescope (La Silla)
during 21 nights between November 10$^{th}$
2007 and January 11$^{th}$ 2008 October
\object{Gl~176} was observed through an I$_c$ 
filter, to maximize the flux of both 
\object{Gl~176} and a M star in the  
11.7' field of view which we planned
to use as photometric reference. That
planned reference however proved 
variable, and we had to fall back
to the average of two fainter
blue stars, with a summed flux of
only 7\% of that of \object{Gl~176}.
In retrospect, this filter choice was therefore 
suboptimal. To minimize atmospheric scintillation
noise we took advantage of the low stellar density to
defocus the images to FWHM $\sim 8\arcsec$, so 
that we could use longer exposure times. The increased 
read-out and sky background noises from the larger 
synthetic aperture which 
we had to use remain negligible compared to both 
stellar photon noise and scintillation.

We gathered 5 to 7 images per night with a
median exposure time of 31~seconds, except 
on December 29$^{th}$ when we obtained sets
of 5 images at three well spaced airmasses to measure
the differential extinction coefficient.  
We tuned the parameters of the \textsc{Iraf Daophot}
package \citep{stetson1987} and optimised the set of 
reference stars to minimise the average dispersion 
in the \object{Gl~176} photometry within the
individual nights. These parameters were then fixed for
the analysis of the full data set. The nightly
light curves for \object{Gl~176} were normalized by
that of the sum of the references, clipped at
3-$\sigma$ to remove a small number of outliers,
and averaged to one measurement per night to examine
the long term photometric variability of \object{Gl~176}.
\object{Gl~176} clearly varies with a $\sim$1.3\% peak-to-peak
amplitude, and a 40-50 days (quasi-)period 
(Fig.~\ref{fig:phot1}). To verify that this variability
does not actually originate in one of the reference stars,
we repeated the analysis using each of the two
reference stars. Those alternate light curves are very
similar to Fig.~\ref{fig:phot1}. The variations are
fully consistent with the 38.92~days period identified 
by \citet{Kiraga2007} in a much longer photometric 
timeseries. Our photometry demonstrates that 
\object{Gl~176}, which \citet{Kiraga2007} find did 
not significantly vary until JD=2453300, has remained 
strongly spotted until the end of our radial velocity 
measurements. 
Our dense sampling also excludes that 38.92~days would 
have been an alias of the true period. We adopt the better
defined \citet{Kiraga2007} value as the rotation period of
\object{Gl~176}.

Our photometric observations are consistent with the signal of
a single spot, within the limitations of their incomplete phase
coverage: the variations are approximately sinusoidal,
and their $\sim$0.2-0.3 phase shift from the corresponding
radial velocity signal closely matches the difference expected
for a spot.
The spot would cover 2.6\% 
of the stellar surface if completely dark,
corresponding to a $\sim 0.16 R_\star$ radius for a circular spot.

\begin{figure*}
\centering
\includegraphics[width=0.9\linewidth]{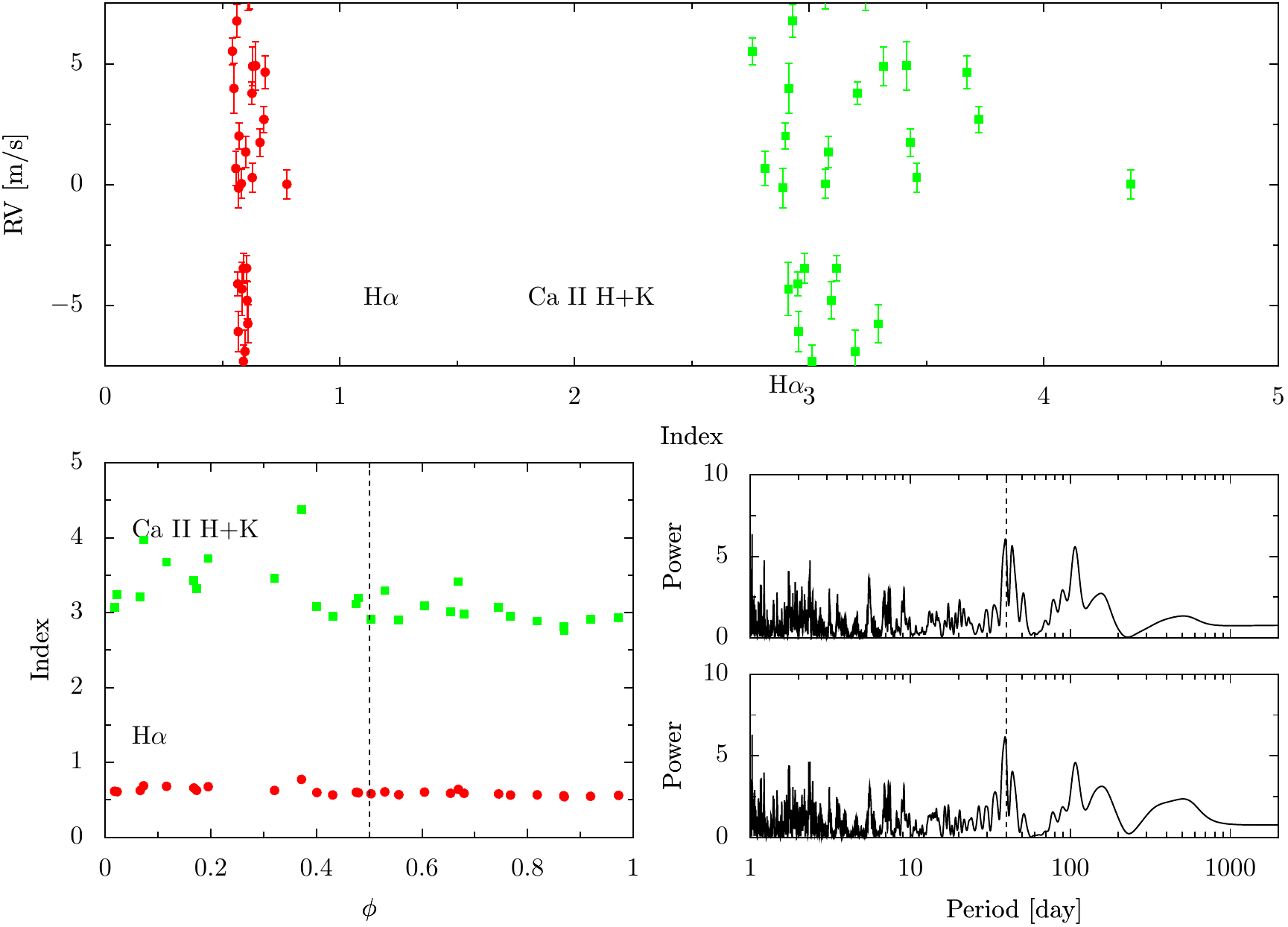}
       \caption{{\it Upper panel:} Differential radial velocity 
     of Gl~176, corrected for the signature of the 8.8~days planet
     in our 2-planet fit, as a function of the H$\alpha$ 
     (red filled circles) and \ion{Ca}{ii} H\&K (green filled squares)
     spectral indices defined in the text for the 2007/2008 observing
     season. 
       {\it Bottom right panels:} the \ion{Ca}{ii} H+K and H$\alpha$ 
     indexes phased to the longer period of the 2-planet 
     model. 
{\it Bottom left panels:} Power Density spectra of the spectroscopic
     indices. A clear power excess peaks at 40 days (vertical 
     dashed lines) 
       }
       \label{fig_index}
\end{figure*}

\subsection{\label{subsect:spec}Variability of the spectroscopic indices}
The emission reversal in the core of the \ion{Ca}{ii} H\&K resonant
lines and in the H$\alpha$ line results from non-radiative heating 
of the chromosphere, which is magnetically coupled to the photospheric 
spots and plages. We measured in the HARPS spectra the spectral 
indices defined by \citet{Bonfils2007} to probe these chromospheric 
spectral features, and examine here their variability.


The power spectra for both the H$+$K and H$\alpha$ 
indices have clear peaks near 40~days 
(Fig.~\ref{fig_index}, lower right panel). 
Within the combined uncertainties, these peaks are 
consistent with both the photometric 
period and the longer radial velocity period. The phasing of
the chromospheric index and the photometry is such
that lower photometric flux matches
higher \ion{Ca}{ii} emission, as expected if active
chromospheric regions hover above dark 
photospheric spots. 

Though certainly not as clearly as for \object{Gl~674} 
\citep{Bonfils2007}, a plot of the (apparent) radial-velocity 
(after subtraction of the 8.8-day planet) against the  
H$+$K spectral index (Fig.~\ref{fig_index}, upper panel) 
similarly suggests the
loop pattern which is expected for a spot
\citep{Bonfils2007}: a spot produces maximal
velocity offsets when it is on either edge of the 
star, where geometric projection reduces the apparent
area of its associated chromospheric emission to
an intermediate value; it produces no velocity
offset when it crosses the sub-observer meridian,
with a maximal projected area for a front-facing crossing 
and a minimal (null for a non-polar spot) projected 
area for a back facing crossing. The radial velocity
offset therefore cancels for both the minimum and the
maximum chromospheric emission, and is maximal for 
intermediate chromospheric emission levels. The
pattern here is definitely noisier than observed
on \object{Gl~674}, suggesting that the spot pattern
may evolve on a time scale of the order of our
observing period.
%

\subsection{\label{sect:blah}Planets vs. activity}

In \S\ref{sect:orbit} we showed that our 57
radial-velocity measurements of \object{Gl~176} are
well described by two Keplerian signals.
Section~\ref{sect:activity} however demonstrates 
that the rotation period of \object{Gl~176} 
coincides with the longer of these two Keplerian 
periods. The stellar flux and the \ion{Ca}{ii} H$+$K 
emission vary with that period, with a phase relative
to the velocity variations consistent with a magnetic
spot on the stellar surface. As a consequence,
some, and probably all, of the 40-day radial-velocity signal
must originate in the spot. Planet-induced activity
through magnetic coupling \citep[e.g.~][]{Shkolnik2005}
would in principle be an alternative explanation of
the correlation, but has never been observed for such
a long-period planet. The inner planet in addition
is not very much less massive than the hypothetical 40-day 
planet. One would, at least naively, expect its 
position in the inner magnetosphere of \object{Gl 176}
to more than make up for its lower mass. The
8.8-day period however is only seen in the radial velocity
signal, and it has no photometric or chromospheric
counterpart.

\section{Discussion and conclusions}
The most important result of the above analysis 
is that a M~sin(i)=8.4~M$_{Earth}$ planet orbits 
\object{Gl~176} in a $\sim$8.8-day orbit. Variability identifies 
the stellar rotation period as 38.92~days, and 
the 8.8-day period therefore cannot reflect rotation 
modulation. The short period signal, in spite of its 
similar amplitude, also has no counterpart
in either photometry or chromospheric emission, further
excluding a signal caused by magnetic activity.

Like  \object{Gl~674}
\citep{Bonfils2007}, \object{Gl~176} demonstrates
that single planets can be identified around
moderately active M-dwarfs, at the cost of
doubling or tripling the number of measurements
over a magnetically quiet M-dwarf.
Since the Keplerian model does not reflect a physical 
reality for the 40-day period, its residuals
must be interpreted with caution. They are well 
above the measurement errors ($\bar{\chi}^2 = 5.86$ 
per degree of freedom), and could in
principle reflect addition planet(s) in the 
system. More likely, much of these residuals
stem from long term evolution of the spot
pattern of \object{Gl~176}. Many additional 
radial velocity measurements would be needed 
to firmly identify additional planets amongst
this spot-evolution noise. That cost may, in practice 
if not in theory, effectively impede the detection of
multi-planet systems around moderately active stars.
It may therefore not be fully by coincidence that 
\object{Gl~674} and \object{Gl~176} are both the 
only M-dwarf planetary systems with no indications
of further planets and the two most active M-dwarfs 
with known planets. 

At 0.066 AU from its parent star, the thermal
equilibrium temperature of \object{Gl~176~b} is 
$\sim$450~K. 
Its  8.4~M$_{Earth}$ M~sin(i) might be sufficient 
for accretion of a significant gas envelope to have
occured, in  particular in case the inclination 
would turn out to be non-trivial, but the rocky 
core most likely dominate the total mass
\citep[e.g.][]{Seager2007,Valencia2007}.

With a mass of only  M~sin(i)=8.4~M$_{Earth}$, 
Gl~176b adds to the growing evidence 
\citep[e.g.][]{Bonfils2007}
that super-Earths are common around very low mass 
stars: 6 of the 20 known planets with 
M~sin(i)~$<$~0.1~M$_{Jup}$ orbit an M dwarf, in contrast 
to just 3 of the $\sim$250 known Jupiter-mass planets.

\begin{acknowledgements}

We would like to thank the ESO La Silla staff for their excellent support
and our collaborators of the HARPS consortium for making this instrument 
such a success, as well as for contributing some of the observations. 
Financial support from the "Programme National de 
Plan\'etologie'' (PNP) of CNRS/INSU, France, is gratefully acknowledged.
XB acknowledge support from the Fundação para a Ciência e a Tecnologia
(Portugal) in the form of a fellowship (reference
SFRH/BPD/21710/2005) and a program (reference
PTDC/CTE-AST/72685/2006), as well as the Gulbenkian Foundation for
funding through the ``Programa de Estímulo à Investigação''.
N.C.S. would like to thank the support from Funda\c{c}\~ao para a Ci\^encia
e a Tecnologia, Portugal, in the form of a grant (references
POCI/CTE-AST/56453/2004 and PPCDT/CTE-AST/56453/2004), and
through programme Ci\^encia\,2007 (C2007-CAUP-FCT/136/2006).

\end{acknowledgements}


\bibliographystyle{aa}
\bibliography{biblio}

\begin{table}
\caption{Radial-velocity measurements and error bars for Gl\,176. 
All values are relative to the solar system barycenter, and
corrected from the small perspective acceleration using the Hipparcos
parallax and proper motion. Only available electronically.}
\label{TableRV}
\centering
\begin{tabular}{c c c}
\hline\hline
\bf JD-2400000 & \bf RV & \bf Uncertainty \\
 & \bf [km\,s$^{-1}$] & \bf [km\,s$^{-1}$] \\
\hline
52986.713028 & 26.4097 & 0.0024 \\
53336.797232 & 26.4133 & 0.0018 \\
53367.703446 & 26.4080 & 0.0010 \\
53371.679444 & 26.4146 & 0.0012 \\
53372.672289 & 26.4150 & 0.0011 \\
53373.698683 & 26.4149 & 0.0016 \\
53375.708263 & 26.4074 & 0.0011 \\
53376.644426 & 26.4074 & 0.0011 \\
53377.637888 & 26.4072 & 0.0010 \\
53378.667446 & 26.4114 & 0.0013 \\ 
53674.790011 & 26.4012 & 0.0010 \\
53693.724506 & 26.4120 & 0.0011 \\
53695.679077 & 26.4167 & 0.0009 \\
53697.762057 & 26.4217 & 0.0016 \\
53699.629044 & 26.4154 & 0.0011 \\
53721.725478 & 26.4120 & 0.0014 \\
53725.600014 & 26.4082 & 0.0014 \\
53727.617518 & 26.4069 & 0.0012 \\
53784.533236 & 26.4056 & 0.0011 \\
53786.526663 & 26.4068 & 0.0010 \\
53809.529447 & 26.4176 & 0.0011 \\
53810.515057 & 26.4184 & 0.0010 \\
53811.510284 & 26.4211 & 0.0011 \\
53812.506114 & 26.4200 & 0.0013 \\
53813.507893 & 26.4148 & 0.0011 \\
53814.507265 & 26.4123 & 0.0011 \\
53815.501823 & 26.4106 & 0.0012 \\
53817.502490 & 26.4110 & 0.0010 \\
54048.826783 & 26.4184 & 0.0010 \\
54050.768921 & 26.4114 & 0.0009 \\
54052.748970 & 26.4067 & 0.0011 \\
54054.812777 & 26.4093 & 0.0010 \\
54078.698716 & 26.4005 & 0.0009 \\
54080.713033 & 26.4065 & 0.0012 \\
54082.712795 & 26.4123 & 0.0012 \\
54084.737341 & 26.4136 & 0.0014 \\
54114.597344 & 26.4051 & 0.0014 \\
54117.631291 & 26.4122 & 0.0010 \\
54122.584109 & 26.4097 & 0.0010 \\
54135.548955 & 26.4141 & 0.0010 \\
54140.552388 & 26.4040 & 0.0010 \\
54142.585254 & 26.4095 & 0.0010 \\
54166.508802 & 26.4099 & 0.0011 \\
54168.505999 & 26.4141 & 0.0010 \\
54170.501820 & 26.4194 & 0.0011 \\
54174.499424 & 26.4092 & 0.0012 \\
54342.888384 & 26.4068 & 0.0011 \\
54345.866037 & 26.4087 & 0.0010 \\
54385.842984 & 26.3989 & 0.0013 \\
54386.803603 & 26.4015 & 0.0015 \\
54387.840396 & 26.4044 & 0.0012 \\
54390.838236 & 26.4087 & 0.0012 \\
54392.803574 & 26.4030 & 0.0011 \\
54393.820127 & 26.4044 & 0.0011 \\
54394.817280 & 26.4029 & 0.0012 \\
54423.739978 & 26.4045 & 0.0013 \\
54428.729841 & 26.4096 & 0.0010 \\
\hline
\end{tabular}
\end{table}

\end{document}